\documentclass{PoS}

\def\tr{{\rm Tr}}

\def\bea{\begin{eqnarray}}
\def\eea{\end{eqnarray}}
\def\nn{\nonumber}
\def\half{\frac{1}{2}}

\def\lmatrix{\left(\begin{array}}
\def\rmatrix{\end{array}\right)}

\def\msbar{\overline{\rm MS\kern-0.5pt}\kern0.5pt}
\def\phat{{\hat p}}
\def\ptilde{{\tilde p}}

\title{The lattice gradient flow at tree level}

\ShortTitle{The lattice gradient flow at tree level}

\author{Zolt\'an Fodor\\
       University of Wuppertal, Department of Physics, Wuppertal D-42097, Germany\\
       J\"ulich Supercomputing Center, Forschungszentrum J\"ulich, J\"ulich D-52425, Germany\\
       E\"otv\"os University, Institute for Theoretical Physics, Budapest 1117, Hungary\\
       \email{fodor@bodri.elte.hu}}

\author{Kieran Holland\\
        University of the Pacific, 3601 Pacific Ave, Stockton CA 95211, USA\\
        Albert Einstein Center for Fundamental Physics, Bern University, Bern, Switzerland\\
        \email{kholland@pacific.edu}}

\author{Julius Kuti\\
        University of California, San Diego, 9500 Gilman Drive, La Jolla, CA 92093, USA\\
        \email{jkuti@ucsd.edu}}

\author{Santanu Mondal\\
        E\"otv\"os University, P\'azm\'any P\'eter s\'et\'any 1, 1117 Budapest, Hungary\\
        MTA-ELTE Lendulet Lattice Gauge Theory Research Group, 1117 Budapest, Hungary\\
        \email{santanu@bodri.elte.hu}}

\author{\speaker{D\'aniel N\'ogr\'adi}\\
        E\"otv\"os University, P\'azm\'any P\'eter s\'et\'any 1, 1117 Budapest, Hungary\\
        MTA-ELTE Lendulet Lattice Gauge Theory Research Group, 1117 Budapest, Hungary\\
        \email{nogradi@bodri.elte.hu}}

\author{Chik Him Wong\\
        University of California, San Diego, 9500 Gilman Drive, La Jolla, CA 92093, USA\\
        \email{rickywong@physics.ucsd.edu}}

\abstract{The cut-off effects of the lattice gradient flow -- often called Wilson flow
-- are calculated on a periodic 4-torus at leading order in the gauge coupling. 
A large class of discretizations is considered which includes
all frequently used cases in practice. It is shown how the results lead to a smoother continuum extrapolation for the
$\beta$-function of $SU(3)$ gauge theory with $N_f = 4$ flavors of fermions.}

\FullConference{The 32nd International Symposium on Lattice Field Theory - Lattice 2014,\\
		June 23-28, 2014\\
		Columbia University New York, NY}

\begin{document}

\section{Introduction}
\label{introduction}

It is well established these days that the Yang-Mills gradient flow, or its lattice implementation the Wilson flow 
\cite{Luscher:2009eq, Luscher:2010iy, Luscher:2010we, Luscher:2011bx}, is
a very useful tool in the non-perturbative study of non-abelian gauge theories. 
Lattice regularization
introduces cut-off dependence which ultimately must be removed to produce
continuum results, e.g. \cite{Borsanyi:2012zs}.

As shown for other observables in QCD, for example quark number susceptibilities, tree-level improvement can lead to a
smoother continuum extrapolation than working directly with the original, unimproved quantities. The first systematic
investigation of tree-level improvement of the observable $E(t)$ (see section \ref{flow} for our definitions) was given
in \cite{Fodor:2014cpa} for a large class of discretizations. The necessary leading order 
lattice perturbation theory calculations were performed
in two setups. First, on a periodic 4-torus of finite volume the necessary formulae are given which include all orders
in $a^2/t$. These may be numerically evaluated to arbitrary precision and can then be used to tree-level 
improve data from simulations. Second, the infinite volume limit is taken and the results are expanded in 
$a^2/t$ leading to optimal simulation parameters.  The latter analytical calculation is checked against the 
former numerical evaluation.
In this contribution we briefly review the results of the infinite volume expansion to order $O(a^8)$
and the finite volume results valid to all orders in $a^2/t$. 

Section \ref{flow} summarizes the necessary details of the
gradient flow,  section \ref{finite} contains the finite volume discussion including the gauge zero mode,
in section \ref{infinite} the infinite volume limit is taken and the results of the expansion in the lattice spacing are
reviewed. We end in section \ref{conc} with a number of conclusions and future directions.

\section{Gradient flow}
\label{flow}

The Yang-Mills gradient flow \cite{Luscher:2009eq,Luscher:2010iy,Luscher:2010we} is an evolution in the space of gauge
fields given by the gradient of the pure gauge action $S_{\rm YM}$,
\bea
\label{floweq}
\frac{d A_\mu(t)}{dt} = - \frac{\delta S_{\rm YM}}{\delta A_\mu}\;,
\eea
where $t$ is the flow time. In a quantum field theory setup this construction can be thought of as defining 
a complicated observable in
the sense that the path integral is over the initial conditions at $t=0$ while the observables are evaluated at $t>0$.
Clearly, the gauge field at $t>0$ is a non-linear complicated function of the initial condition at $t=0$. As shown in  
\cite{Luscher:2009eq,Luscher:2010iy,Luscher:2010we} the flow is essentially a smoothing operation over the range of
$\sqrt{8t}$ and leads to $UV$-finite composite operators in some cases where the original composite operator was
$UV$-divergent. For example the composite operator $E = - \half \tr F_{\mu\nu} F_{\mu\nu}$ at $t>0$ is finite in perturbation
theory provided the gauge coupling is renormalized in the usual way, for example in $\msbar$,
\bea
\langle t^2 E(t) \rangle = \frac{3(N^2-1)g^2}{128\pi^2}\left( 1 + O(g^2) \right)\;.
\label{key}
\eea
The above relation can be turned around and the quantity $E(t)$ can be used to define a renormalized coupling scheme.
The scale of the running in infinite volume is $\mu = 1/\sqrt{8t}$. The size of cut-off effects in this setup is
detailed in section \ref{infinite}.

The calculation of the running coupling on the lattice can most conveniently be performed by step scaling \cite{Luscher:1991wu}
where the
running scale is the inverse of the linear size of the system $\mu = 1/L$. Using the gradient flow in a step scaling
study necessitates that the new scale $t$ is tied to the size of the box and $\sqrt{8t}/L$ is kept constant. Otherwise
the problem would have two independent scales, $\sqrt{8t}$ and $L$. This is very similar to using Wilson loops for the
definition of the running coupling where the ratio of the size of the Wilson loop and $L$ is kept constant.

The fact that finite volume is involved necessitates to rederive the key equation (\ref{key}) in finite volume
with a given choice of boundary conditions. In \cite{Fodor:2012td,Fodor:2012qh} periodic boundary conditions were chosen
for the gauge field, which we also use here. To leading order the corresponding result is
\bea
\langle t^2 E(t) \rangle = \frac{3(N^2-1)g^2}{128\pi^2}\left(1+\delta\left(\sqrt{8t}/L\right)\right)\;,
\label{keyv}
\eea
where
\bea
\delta(\sqrt{8t}/L) &=& - \frac{64t^2\pi^2}{3L^4} + \vartheta^4\left(\exp\left(-\frac{L^2}{8t}\right)\right) - 1\;,
\eea
using the $3^{rd}$ Jacobi elliptic function $\vartheta$.
In section \ref{finite} the cut-off effects associated with the finite volume setup are given, i.e. the lattice spacing
dependence of the finite volume correction factor $\delta(\sqrt{8t}/L)$ which we will label $\delta(\sqrt{8t}/L,a^2/t)$.
Note that $\sqrt{8t}/L$ is always kept constant.

In leading order lattice perturbation theory the discretization of $E(t)$ is entirely given by the quadratic term of
the lattice gauge action. The gauge action actually enters the calculation in three instances: the action used along the
gradient flow, the dynamical gauge action used for the simulation and finally the discretization of the observable $E$
itself.
In the continuum all three are the same, $- \half \tr F_{\mu\nu} F_{\mu\nu}$, but on the lattice they may be 
different. In this work we consider a broad class of discretizations where the first two instances may be the Symanzik
improved actions and for the third instance we also consider
the symmetric clover discretization along with the Symanzik-improved action. The Symanzik-improved actions may
potentially have different improvement coefficients $c_1$. 
Clearly, all cases used in practice are covered by our class of discretizations.

For the frequently used cases, Wilson plaquette ($c_1 = 0$), tree-level Symanzik ($c_1=-1/12$) or clover,
let us introduce the notation $WWC$, $SSS$, $SSC$, etc, where the order is
always Flow-Action-Observable.

\section{Finite volume}
\label{finite}

The quadratic terms in the action that determine the leading order result are well-known. For the Symanzik-improved
gauge action with improvement coefficient $c$ we have
\bea
\label{syms}
S_{\mu\nu} = \delta_{\mu\nu} \left( \phat^2 - a^2 c \sum_\rho \phat_\rho^4 - a^2 c \phat_\mu^2 \phat^2 \right) 
- \phat_\mu \phat_\nu \left( 1 - a^2 c ( \phat_\mu^2 + \phat_\nu^2 ) \right) \;,
\eea
while for the clover discretization it is \cite{Fritzsch:2013je}
\bea
\label{clov}
K_{\mu\nu} = \left( \delta_{\mu\nu} \ptilde^2 - \ptilde_\mu \ptilde_\nu \right) 
\cos\left( \frac{ap_\mu}{2} \right) \cos\left( \frac{ap_\nu}{2} \right)\;,
\eea
using the lattice momenta
\bea
{\hat p}_\mu = \frac{2}{a} \sin\left( \frac{ap_\mu}{2} \right)\;, \qquad {\tilde p}_\mu = \frac{1}{a} \sin( a p_\mu )\;.
\eea
In order to have a simple notation let us further introduce
\bea
{\cal S}^{f}_{\mu\nu} &=& S_{\mu\nu}(c=c_f) \nn \\
{\cal S}^{g}_{\mu\nu} &=& S_{\mu\nu}(c=c_g)  \\
{\cal S}^{e}_{\mu\nu} &=& S_{\mu\nu}(c=c_e)\;, \qquad {\rm or} \qquad K_{\mu\nu}\;. \nn
\eea
for the specific choices of the three ingredients: the flow, the dynamical action and the observable. For more details
on lattice perturbation theory see \cite{Weisz:1982zw, Symanzik:1983dc, Weisz:1983bn, Luscher:1984xn, Luscher:1985zq}.

\begin{figure}
\begin{center}
\includegraphics[width=7.5cm]{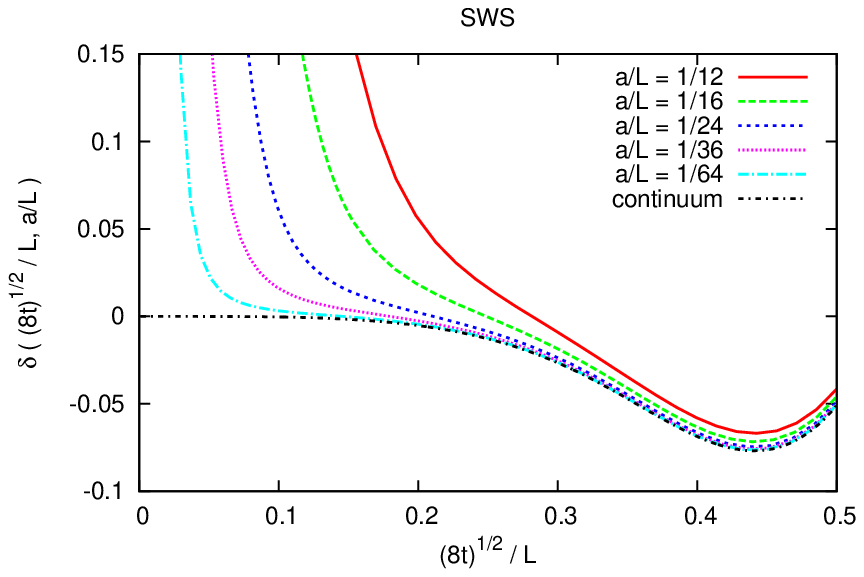}  \includegraphics[width=7.5cm]{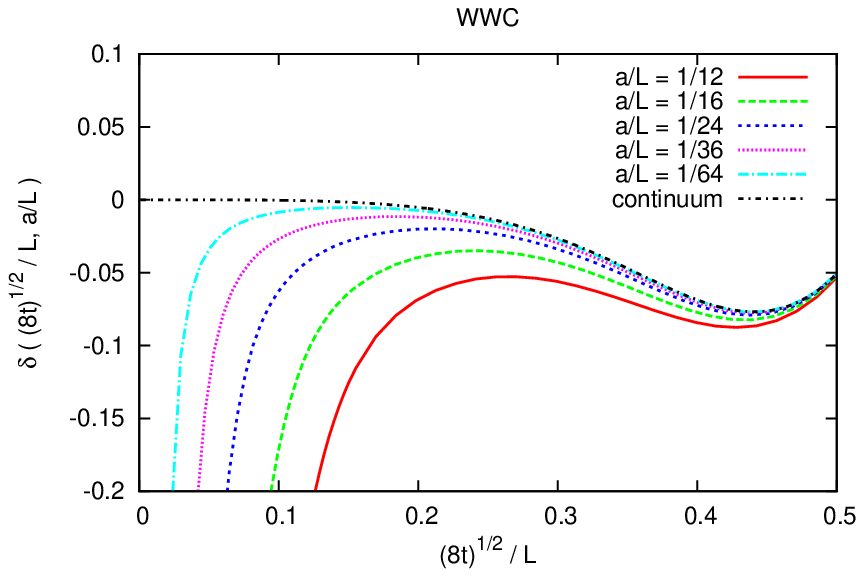} \\
\includegraphics[width=7.5cm]{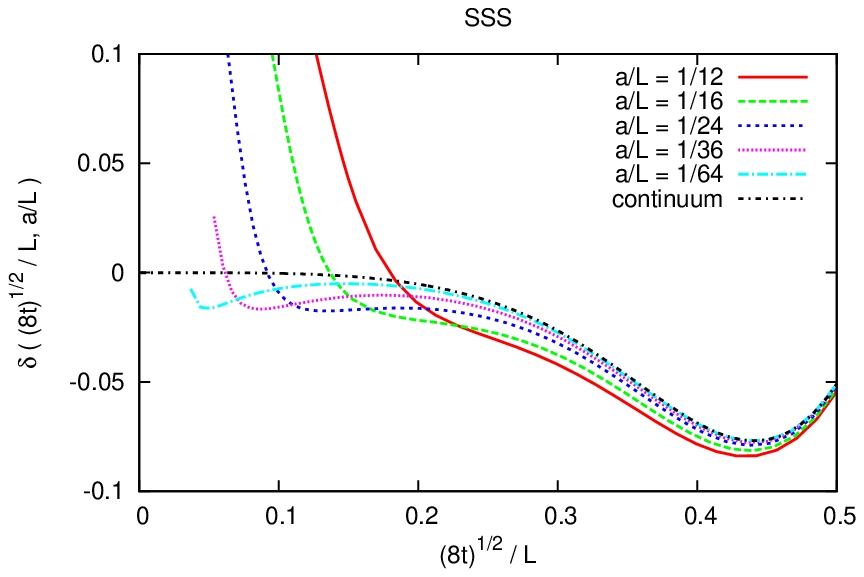}  \includegraphics[width=7.5cm]{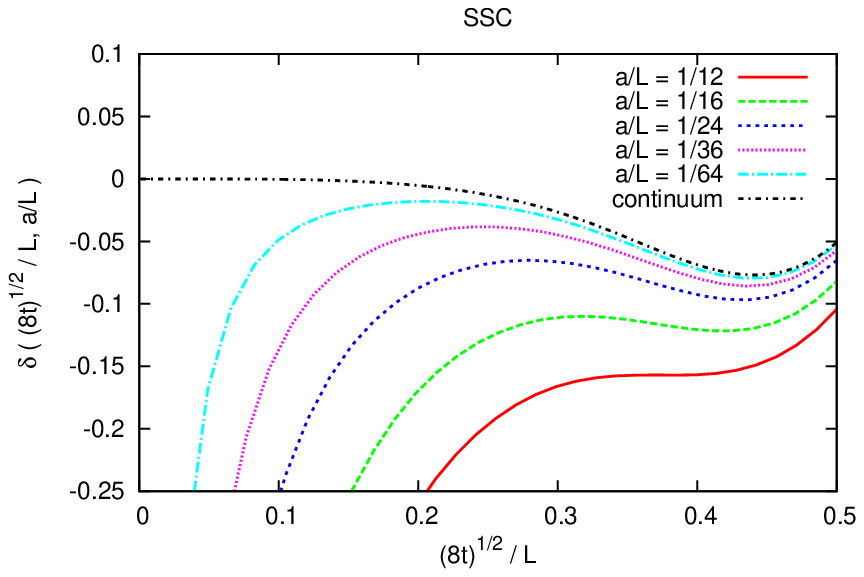} 
\end{center}
\caption{The tree-level finite volume and finite lattice spacing correction factors
$\delta(\sqrt{8t}/L, a/L) = C(a^2/t,\sqrt{8t}/L) - 1$
for four examples, the $SWS$, $WWC$, $SSS$ and $SSC$ cases as a function of $\sqrt{8t}/L$ at various lattice spacings.
The continuum result is from \cite{Fodor:2012td}.}
\label{delta}
\end{figure}

In a periodic finite volume setting special attention needs to be paid to the gauge zero modes. This problem has been
dealt with in \cite{Fodor:2012td, Fodor:2012qh} in the continuum and it is easy to see that to
leading order in the coupling the contribution of the zero mode at finite lattice
spacing will be the same as in the continuum. The finite volume also means that the momenta are quantized $p_\mu =
2\pi/L \cdot n_\mu$ with integers $n_\mu$ and the finite lattice spacing restricts their range to $0 \leq n_\mu < L/a$. Since
the zero mode is treated separately, $n^2 \neq 0$.

\begin{figure}
\begin{center}
\includegraphics[width=7.5cm]{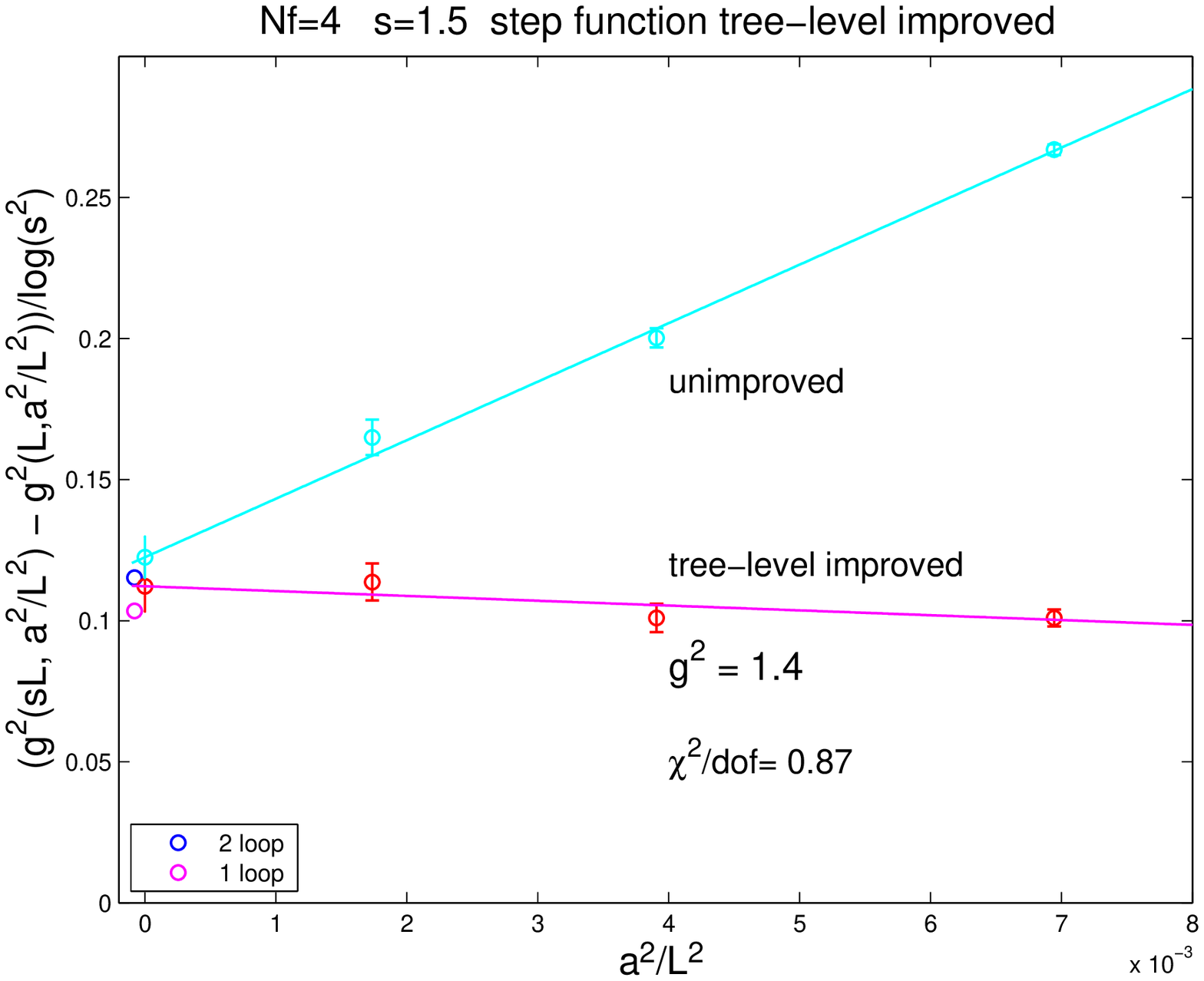} \includegraphics[width=7.5cm]{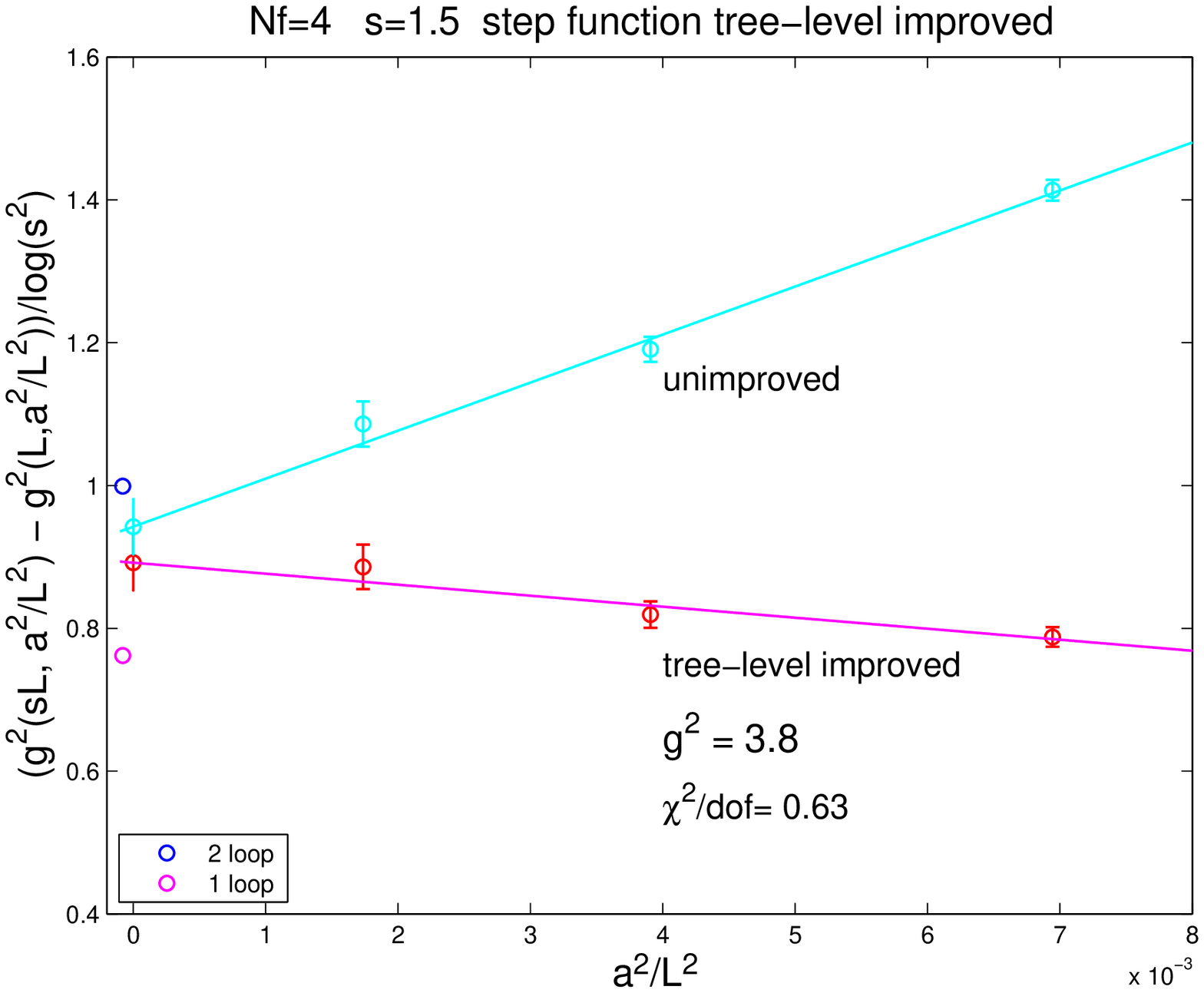}
\end{center}
\caption{Continuum extrapolations of the discrete $\beta$-function for two selected $g^2$ values $1.4$ (left) and
$3.8$ (right) for $N_f = 4$ flavors with and without tree-level improvement. The data is from \cite{Fodor:2012td}.}
\label{beta}
\end{figure}

\begin{figure}
\begin{center}
\includegraphics[width=7.5cm]{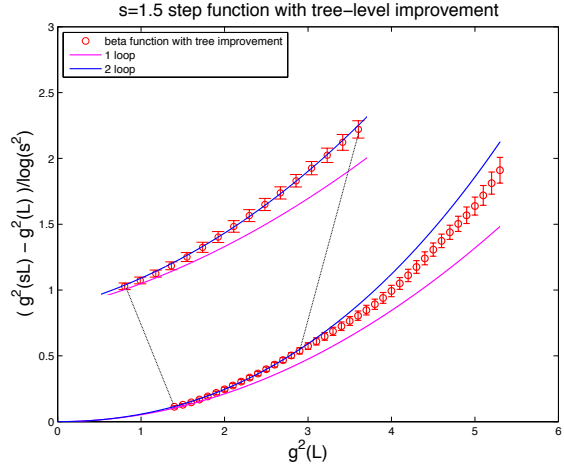}
\end{center}
\caption{Continuum extrapolated discrete $\beta$-function of $N_f = 4$ flavors. A section is magnified for better
visibility, the data is from \cite{Fodor:2012td}.}
\label{beta2}
\end{figure}

Combining all ingredients the leading order result for $C(a^2/t, \sqrt{8t}/L) = 1 + \delta(\sqrt{8t}/L,a^2/t)$ is
\bea
\label{catv}
C(a^2/t,\sqrt{8t}/L) &=& \frac{128\pi^2 t^2}{3L^4} + \frac{64\pi^2 t^2}{3L^4} 
\sum_{n_\mu = 0,\; n^2\neq 0}^{L/a-1} \tr\, \left( e^{-t\left({\cal S}^f + {\cal G}\right)} ({\cal S}^g + {\cal G})^{-1}
e^{-t\left({\cal S}^f + {\cal G}\right)} {\cal S}^e \right)\;, \nn \\
&& 
\eea
which may be evaluated numerically to arbitrary precision. Here the exponential terms are coming from
the leading order solution of the flow equation (\ref{floweq}), the term with the inverse is the leading order gluon
propagator and the ${\cal S}^e$ term is simply the observable itself. For the sake of convenience a gauge fixing term
${\cal G}_{\mu\nu} = {\hat p}_\mu {\hat p}_\nu / \alpha$ was introduced. In principle the gauge fixing parameter may be different
along the flow and in the propagator but we chose $\alpha = 1$ for both. The final result is of course independent of
these choices. In front of the sum above is the contribution of the zero mode.
The factor $\delta(\sqrt{8t}/L,a^2/t)$ is shown in figure \ref{delta} for various choices of discretizations and various
choices of lattice volume $L/a$. The continuum $\delta(\sqrt{8t}/L)$ factor is also shown for comparison.

In order to test the usefulness of tree-level improvement we consider the $\beta$-function of $SU(3)$ gauge theory with 
$N_f = 4$
flavors \cite{Fodor:2012td}. The discrete $\beta$-function was computed there and here we reanalyze the data
corresponding to two examples of continuum extrapolation, with and without tree-level improvement. The extrapolations
for fixed $g^2 = 1.4$ and $g^2 = 3.8$ are shown in figure \ref{beta} corresponding to a scale change of $s=3/2$. 
Clearly the slope of the extrapolation reduced considerably with tree-level improvement. On figure \ref{beta2} we show
the continuum extrapolated discrete $\beta$-function over the entire range of couplings.

\section{Infinite volume}
\label{infinite}

\begin{figure}
\begin{center}
\includegraphics[width=7.5cm]{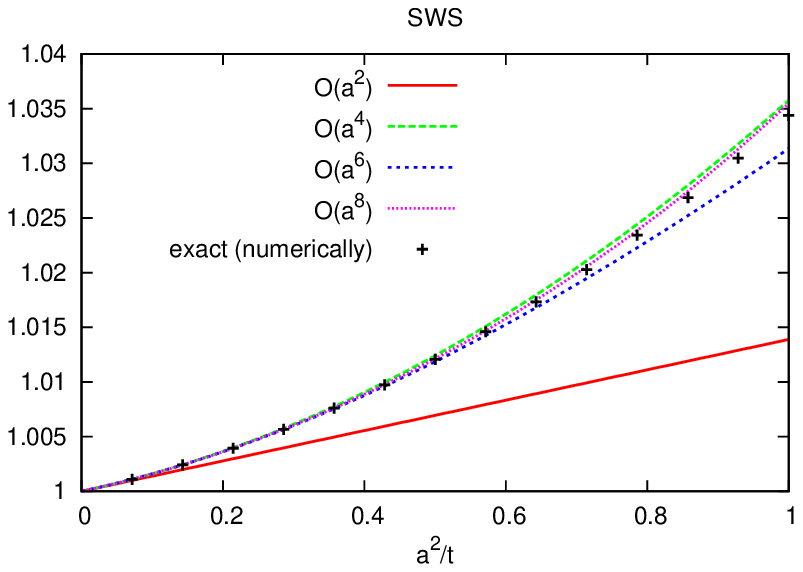}  \includegraphics[width=7.5cm]{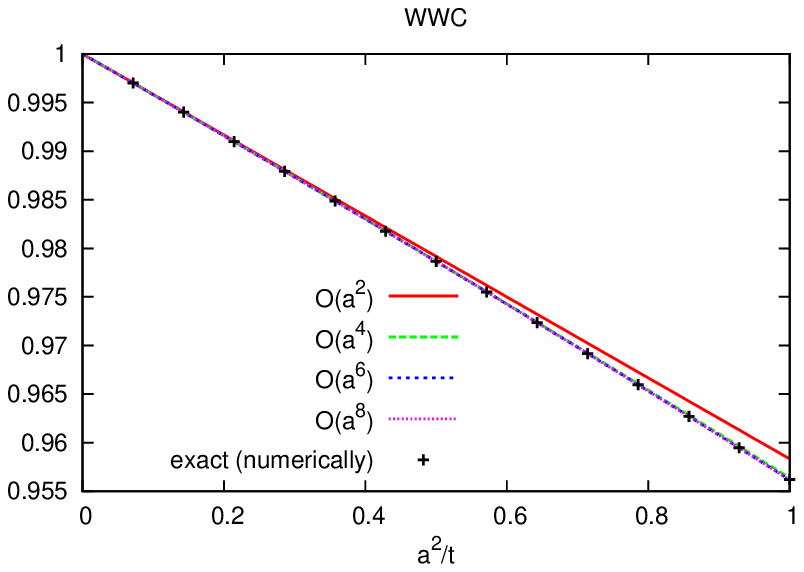} \\
\includegraphics[width=7.5cm]{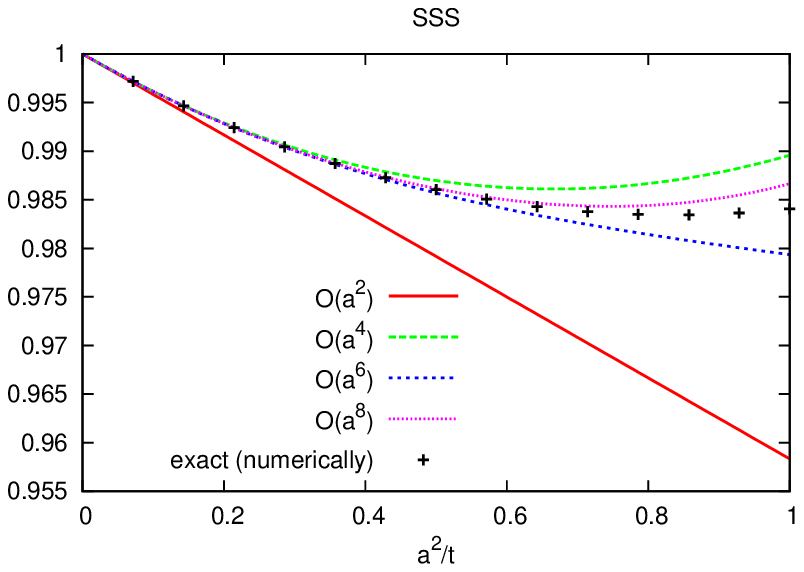}  \includegraphics[width=7.5cm]{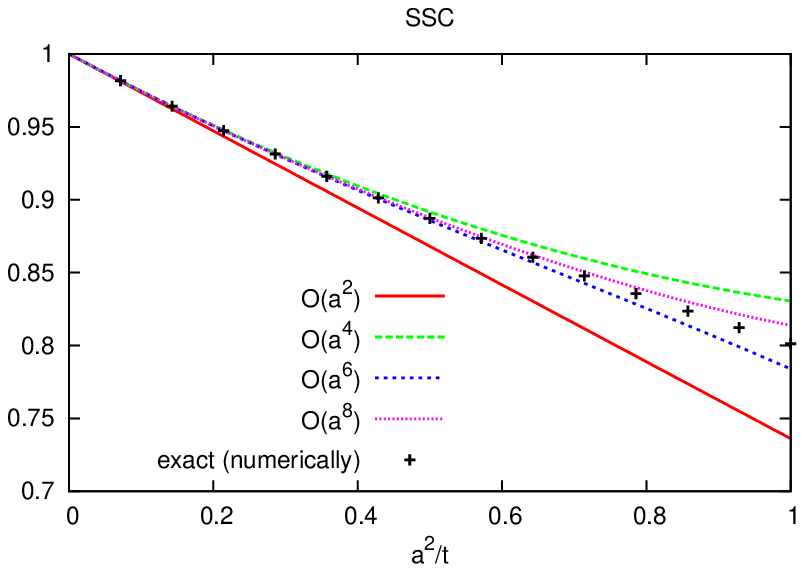} 
\end{center}
\caption{The tree-level improvement factor for four examples, the $SWS$, $WWC$, $SSS$ and $SSC$ cases, 
starting from only the $O(a^2)$ correction and including higher and
higher order terms. The numerical result containing all orders in $a^2/t$, is also shown.}
\label{conv}
\end{figure}

The results in the previous section contained the leading order result in the coupling but all orders in $a^2/t$ or
equivalently $a^2/L^2$, in finite volume. In infinite volume one may obtain analytical results if the expressions are
expanded in $a^2/t$. It is natural to restrict attention to the first few coefficients in an expansion in $a^2/t$ 
because the leading loop effects $g^2 a^2/t$ will certainly be larger than some high power tree-level term 
$a^{2m}/t^m$ for some $m$.

Taking the infinite volume limit of (\ref{catv}) is relatively straightforward. The contribution of the zero modes
disappears and the sum turns into a lattice momentum integral,
\bea
\label{eee}
C(a^2/t) = \frac{64\pi^2t^2}{3} \int_{-\frac{\pi}{a}}^{\frac{\pi}{a}} \frac{d^4p}{(2\pi)^4} 
\tr\, \left( e^{-t\left({\cal S}^f + {\cal G}\right)} ({\cal S}^g + {\cal G})^{-1} e^{-t\left({\cal S}^f + {\cal
G}\right)} {\cal S}^e \right)\;.
\eea
The above expression may again be evaluated numerically to arbitrary precision.
It may also be expanded in $a^2/t$ and the first few coefficients in $C(a^2/t) = 1 + \sum_m C_{2m} (a^2/t)^m$
can be found in \cite{Fodor:2014cpa}.
The relative size of the various orders can be judged by looking at figure \ref{conv} which shows
various truncations and the full, numerically evaluated, result for a few examples.

\section{Conclusion}
\label{conc}

As the application of the gradient flow expands to more areas a systematic understanding of its cut-off
effects becomes essential. In this contribution we reviewed the size of cut-off effects for the
particularly important observable $E(t)$ for a large class of discretizations at tree-level of the coupling. 
The main motivation was to understand cut-off effects in setups frequently used in practice. A natural next step would
be to obtain the corresponding formulae to 1-loop.

\section*{Acknowledgments}

This work was supported by the DOE under grants DOE-FG03-97ER40546 and
DOE-FG02-97ER25308, 
by the Deutsche Forschungsgemeinschaft
grants FO 502/2 and SFB-TR 55 and by the NSF under grants 0704171, 0970137 and 1318220 
and by OTKA under the grant OTKA-NF-104034.

\end{document}